\documentclass[twocolumn,showpacs,preprintnumbers,amsmath,amssymb]{revtex4}

\usepackage{epsf}
\usepackage{graphicx}
\usepackage{dcolumn}
\usepackage{bm}

\begin{document}

\title{NSQUID arrays as conveyers of quantum information}

\author{Qiang Deng and Dmitri V. Averin}

\affiliation{Department of Physics and Astronomy, Stony Brook
University, SUNY, Stony Brook, NY 11794-3800}


\begin{abstract}

We have considered the quantum dynamics of an array of
nSQUIDs -- two-junction SQUIDs with negative mutual inductance
between their two arms. Effective dual-rail structure of
the array creates additional internal degree of freedom for the fluxons in the array, which can be used to encode and transport quantum information. Physically, this degree of freedom is represented by electromagnetic excitations localized on the fluxon. We have calculated the spatial profile and frequency spectrum of these excitations. Their dynamics can be reduced to two quantum states, so that each fluxons moving through the array carries with it a qubit of information. Coherence properties of such a propagating qubits in the nSQUID array are characterized by the dynamic suppression of the low-frequency decoherence due to the motion-induced spreading of the noise spectral density to a larger frequency interval.

\end{abstract}

\date{\today}

\maketitle

\section{Introduction}
Coherence properties and precision of control over the dynamics of superconducting qubits (see, e.g., recent experiments \cite{neeley2010,sun2010,paik2011,shankar2013,ucsb2014}) have reached the level when it becomes possible and interesting to discuss potential architecture of the superconducting quantum computing circuits, either within the gate-model paradigm \cite{nist2007,ucsb2011,fedorov2012,mirrahimi2014} or the adiabatic ground-state approach \cite{dwave2011,bian2013}. Besides the formidable problem of maintaining the level of qubit coherence with increasing circuit complexity, the central issue that needs to be addressed by any architecture of scalable quantum computing devices is the requirement of rapid transfer of quantum information among a large number of qubits with sufficient fidelity. So far, the suggested solutions to the problem of transfer of quantum information were based on controllable direct coupling of qubits or coupling through a common resonator. These solutions, while working nicely for the circuits of few qubits, can not be scaled easily to larger circuits. The purpose of this work is to suggest another approach to the problem of information transfer along a quantum circuit of the superconducting qubits utilizing quantum dynamics of magnetic flux, in which the quantum information is transported along the circuit by propagating classical pulses. This  approach uses the arrays of two-junction SQUIDs, where each of them has a negative mutual inductance between its two arms. Dynamics of such ``nSQUIDs'' \cite{nsquid1} can be represented in terms of the two degrees of freedom, the ``differential'' mode and the ``common'' modes, with very different properties. The former can be used to encode quantum information, while the latter -- to transport it. Then, the overall architecture of a quantum computing circuit built of nSQUIDs is very similar to superconducting classical reversible circuits also based on nSQUIDs \cite{nsquid2,nsquid3}, in which the computation is organized around the information-carrying pulses propagating along the circuit.

Existence the two degrees of freedom with different properties makes nSQUID arrays qualitatively and advantageously different from previously considered arrays of superconducting qubits -- see, e.g., \cite{romito,lyakhov,strauch}, or spins \cite{zwick,bruderer,richert},
as tools for quantum information transfer. In general, a physical variable encoding quantum information and the one used to carry it should satisfy completely different sets of requirements, and in the case of nSQUIDs, the common and the differential dynamic modes can be optimized separately to satisfy these different requirements. Most importantly, the common mode, i.e. the degree of freedom transporting the quantum information does not necessarily need to be itself quantum, since a classical dynamics is sufficient to transport a qubit along the array. This makes potential operation of nSQUID arrays as conveyers of quantum information considerably more straightforward, since it avoids a challenging problem of maintaining quantum coherence of fluxons, supported by the common mode, along a large array of junctions.

\section{Basic model of nSQUID array}

\begin{figure}
\includegraphics[width=4.4cm]{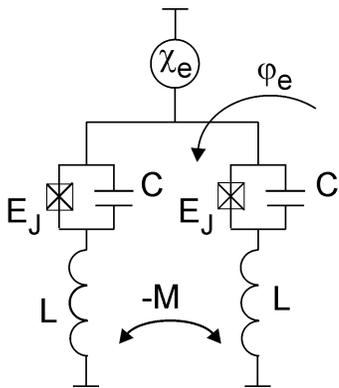}
\caption{Equivalent circuit of an nSQUID: two-junction SQUID with junction capacitances $C$ and Josephson coupling energies $E_J$ and the negative mutual inductance $-M$ between its inductive arms with inductances $L$. The negative mutual inductance makes the effective inductance of the common mode of the SQUID dynamics much smaller than the inductance of the differential mode. Also shown are the phase bias $\chi_e$ of the common mode and $\phi_e$ of the differential mode.}
\label{f1} \end{figure}

We begin our detailed discussion with a description of the elementary cell of the arrays considered in this work -- nSQUID \cite{nsquid1}, a two-junction SQUID with a negative inductance between its two arms (Fig.~\ref{f1}). Dynamics of this structure can be separated naturally into the dynamics of two degrees of freedom: common mode representing the  total current flowing through the two junctions of the SQUID, and the differential mode which represents the difference of the two junction currents, i.e., the current circulating along the SQUID loop. In the configuration of a one-dimensional array, these two degrees of freedom give rise to the two different excitation modes of the array. The common mode corresponds to excitations propagating along  the array and, in the appropriate regime, takes the form of individual fluxons.

The main difference between the nSQUID and the usual two-junction SQUID can be seen if one thinks very crudely about two arms of a SQUID as two parallel wires. For the plain wires, the mutual inductance $M$ between the wires is positive, $M>0$, ensuring  that the effective inductance of the differential mode is always smaller than that of the common mode. In this case, the differential mode can have a non-trivial dynamics only together with the common mode. By contrast, the negative mutual inductance $-M$ between the SQUID arms makes the effective inductance of the differential mode larger than the inductance of the common mode. As a result, one can realize a situation when the differential mode exhibits a non-trivial, e.g. bi-stable, dynamic at low frequencies without exciting the common mode which supports only the excitations with larger frequencies. The differential mode can then be used to encode information, while the dynamics of the common mode is optimized separately for transfer of this information along the array. If the dynamics of both modes is classical, nSQUID structures provide a convenient basis for implementation of classical reversible computing \cite{nsquid2,nsquid3}. If, however, parameters of the differential mode are such that its behavior is quantum, it can be used to encode quantum information, which can then be transported along the array by the evolution of the common mode.

Hamiltonian $H$ of the individual nSQUID (Fig.~\ref{f1}) is given by the standard expression which includes the charging energies and the Josephson coupling energies of the two SQUID junctions. Adding the magnetic energy of the two inductive arms of the SQUID coupled by the negative mutual inductance, one obtains the following expression for the nSQUID Hamiltonian
\begin{eqnarray}
H=\frac{K^2}{2C_t} +\frac{Q^2}{4C}-2E_J\cos \chi \cos \phi
\nonumber \\ + \Big(\frac{\Phi_0}{2\pi}\Big)^2 \Big[\frac{(\chi -
\chi_e)^2}{L-M} + \frac{(\phi - \phi_e)^2}{L+M} \Big] \, .
\label{e1} \end{eqnarray}
Here $\Phi_0=\pi\hbar/e$ is the magnetic flux quantum, $K$ and
$\chi$ are the variables of the common mode: $K=Q_1+Q_2$ is the
total charge on the two capacitances of the SQUID junctions, and
$\chi=(\phi_1+\phi_2)/2$ is the average Josephson phase difference
across the junctions. The common mode has effective inductance $(L-M)/2$ and capacitance $C_t$, which in the case of the circuit in Fig.~\ref{f1}, is equal to the total capacitance $2C$ of the two junctions, but in general can include additional contributions from the external biasing circuit (as, e.g., in Fig.~\ref{f2} below). In quantum dynamic, $K$ and $\chi$ are canonically conjugated variables that satisfy the commutation relation $[\chi,K]=2ei$, standard for the charge and phase of a Josephson junction. The corresponding variables of the differential mode are the charge difference $Q=Q_1-Q_2$ and the phase difference $\phi=(\phi_1-\phi_2)/2$ which have the same commutation relation $[\phi,Q]=2ei$. The effective inductance and capacitance of this mode are $2(L+M)$ and $C/2$ respectively. (Their apparent values in Eq.~(\ref{e1}) are different because of the chosen  normalization of $Q$ and $\phi$.)

As mentioned above, qualitative effect of the negative mutual inductance is to make the dynamic properties of the common and the differential mode in the Hamiltonian (\ref{e1}) of one nSQUID very different. If one neglects for a moment the Josephson coupling, the resonance frequencies of the two modes are $[2/(L-M)C_t]^{1/2}$ and  $1/[(L+M)C]^{1/2} \equiv \omega_p$, and in the large-negative-inductance limit, $M\rightarrow L$, the excitation frequency of the common mode becomes much larger than that of the differential mode, making it possible to clearly separate the frequency ranges of the dynamics of the two modes. As a result, when the nSQUID cells are connected in an array as in Fig.~\ref{f2}, the two modes can be used to perform different functions. In particular, if the coupling inductances $L_C$ are designed to have negative mutual inductance $-M_C$ between them, the main feature of the nSQUID dynamics is preserved: the common mode remains rigid, i.e., it is not affected by the evolution of the differential mode, and is essentially fixed at some value which is either applied externally or generated dynamically. This phase plays then the role of the qubit control signal which is distributed along the array through the ``clock'' line (upper horizontal line in Fig.~\ref{f2}). The differential mode can be used to encode a classical or quantum bit of information in the current circulating along the coupled SQUID loops. Dynamics of the common mode ensures then that the information encoded by the differential mode is transported along the array.

Quantitatively, we consider the arrays with no bias phases applied externally either for the common or differential mode; rather the bias phases $\chi_{e,j}$ for the common modes $\chi_j$ of the array cells, where the index $j$ numbers the cells, are generated dynamically by the array structure. The phase $\chi_{e,j}$ is the phase of the superconducting order parameter at the bias node point of the $j$th nSQUID, i.e., the points connected by inductance $L_0$ in Fig.~\ref{f2}. The Hamiltonian $H_0$ of such an array, a segment of which is shown in Fig.~\ref{f2}, can be written as
\begin{eqnarray}
H_0=\sum_j \Big\{ H^{(j)} + \Big(\frac{\hbar}{2e}\Big)^2 \Big[ \frac{(\chi_{e,j}-\chi_{e,j-1})^2}{2L_0} \nonumber \\ +\frac{(\phi_j - \phi_{j-1})^2}{L_C+M_C} +\frac{(\chi_{e,j}-\chi_j + \chi_{j-1} - \chi_{e,j-1})^2}{L_C-M_C} \Big] \Big\} \, ,  \label{e2a} \end{eqnarray} %
where the sum runs over all nSQUID cells of the array. Each term $H^{(j)}$ in this expression,
\begin{eqnarray}
H^{(j)} = \frac{K_j^2}{4C} +\frac{Q_j^2}{4C}+\frac{K_{e,j}^2}{2C_0} +\Big(\frac{\hbar}{2e}\Big)^2 \Big[\frac{(\chi_{e,j} - \chi_j)^2 }{L-M} \nonumber \\ + \frac{\phi_j^2}{L+M} \Big]- 2E_J\cos \chi_j \cos \phi_j  \, , \label{e2b}
\end{eqnarray}
is the Hamiltonian of the $j$th cell, while the rest of the terms in Eq.~(\ref{e2a}) describe the coupling between the nearest-neighbor cells due to inductances $L_0$ and $L_C$. In addition to the Hamiltonian (\ref{e1}), each term $H^{(j)}$ includes now the charging energy of the charge $K_{e,j}$ on the capacitance $C_0$ of the $j$th cell. The charge $K_{e,j}$ is the conjugate variable to the phase $\chi_{e,j}$, with the standard commutation relation  $[\chi_{e,j},K_{e,j}]=2ei$. As shown explicitly in the Appendix, in the limit of strong negative coupling of the array inductances, $M\rightarrow L$, $M_C\rightarrow L_C$, when the characteristic features of the nSQUID dynamics manifest themselves most prominently, the common phase of each cell of the array becomes pinned down to the corresponding bias phase $\chi_{e,j}$, $\chi_j\simeq \chi_{e,j}$, and the array Hamiltonian $H_0$ reduces to
\begin{eqnarray}
H_0=\sum_j \Big\{ \frac{K_j^2}{2C_t}+\frac{Q_j^2}{4C} - 2E_J\cos \chi_j \cos \phi_j \nonumber \\ +\Big(\frac{\hbar}{2e}\Big)^2 \Big[ \frac{(\chi_{j}-\chi_{j-1})^2}{2L_0} + \frac{\phi_j^2}{2L}+ \frac{(\phi_j - \phi_{j-1})^2}{2L_C} \Big] \Big\} \, ,
\label{e3a} \end{eqnarray}
where $C_t=2C+C_0$. Note that although we included for uniformity the negative coupling $M_C$ in the requirements of the strong-negative-coupling limit, in principle, as follows from the  discussion in the Appendix -- cf. Eq.~(\ref{a7}), the condition $M\rightarrow L$ alone is sufficient for the reduction of the array Hamiltonian to form (\ref{e3a}) with different effective coupling inductance.

\begin{figure}
\includegraphics[width=6.4cm]{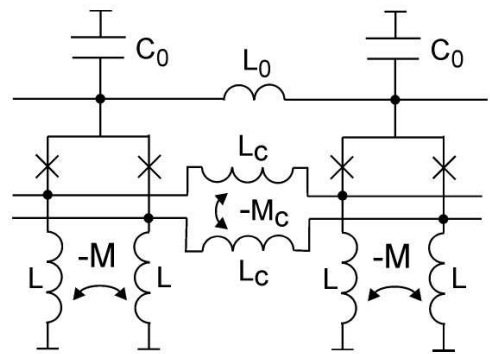}
\caption{Dual-rail Josephson array made of nSQUID cells shown in
\protect Fig.~\ref{f1}. The array cells are coupled by inductances $L_C$, with negative mutual inductance $-M_C$ between them. No bias phases are applied externally either for the common or differential mode; the bias for the common mode is generated self-consistently by the array dynamics. In this dynamics, the common mode plays the role of the qubit control signal propagating along the control line with specific capacitance $C_0$ and inductance $L_0$, whereas the differential mode encodes a qubit of quantum information that is being transported along the array. } \label{f2}
\end{figure}

In what follows, we consider the situation when both the clock phases $\chi_j$, and the information phases $\phi_j$ can be described in the continuous approximation,  $\chi_j, \phi_j \rightarrow \chi(x), \phi(x)$, where $x$ is a continuous dimensionless position along the array (which can be understood as the real spatial coordinate in units of the size of the elementary cell of the array).
Such an approximation is strictly valid if the characteristic length of variation of each phase, ``Josephson penetration length'', is large, $\lambda_0,\lambda_C\gg 1$, where
\[ \lambda_0\equiv (\hbar/2e)/[2 E_J L_0]^{1/2}, \;\;\; \lambda_C \equiv (\hbar/2e)/ [2 E_J L_C]^{1/2}. \]
In the continuous approximation, the Hamiltonian (\ref{e3a}) of the array can be written as
\begin{eqnarray}
H_0 =\int dx \Big\{ \Big(\frac{\hbar}{2e}\Big)^2 \Big[C_t \dot{\chi}^2+ \frac{(\chi')^2}{2L_0}+C\dot{\phi}^2 + \frac{\phi^2}{2L} \nonumber \\ + \frac{(\phi')^2}{2L_C}\Big] -2E_J\cos \chi(x,t) \cos \phi(x,t) \Big\} \, ,
\label{e3} \end{eqnarray}
where the prime denotes the derivative with respect to $x$, and all the parameters, $C_t,L_0,C,L_C,E_J$, are defined now per unit length, with the exception of inductance $L$, for which the inverse inductance is proportional to length, and one defines $1/L$ per unit length.

In the situation of interest for quantum information transfer, we can also adopt an assumption that the dynamics of $\phi$ is restricted to the regime of small phases, $|\phi| \ll 1$, since the qubit designs aim  typically at this regime to minimize the decoherence effects. In this case, one can expand the Hamiltonian (\ref{e3}) in $\phi$ and express it as a sum of the parts governing the evolution of the clock and information phases:
\[H_0=H^{(\chi)}+H^{(\phi)}, \]
where
\begin{equation}
H^{(\chi)} =\int dx \Big\{ \Big(\frac{\hbar}{2e}\Big)^2 \Big[ \frac{C_t \dot{\chi}^2}{2} + \frac{(\chi')^2}{2L_0}\Big] -2E_J\cos \chi \Big\} \, ,
\label{e4a} \end{equation}
and, in the quadratic approximation,
\begin{eqnarray}
H^{(\phi)} =\int dx  \Big\{ \Big(\frac{\hbar}{2e}\Big)^2 \Big[C\dot{\phi}^2 + \frac{(\phi')^2}{2L_C} +\frac{\phi^2}{2L}\Big]
\nonumber \\ + E_J\cos \chi(x,t) \phi^2\Big\} \, .
\label{e4b} \end{eqnarray}

The $\chi$-part (\ref{e4a}) is equivalent to the sine-Gordon Hamiltonian of a regular long Josephson junction. For the purpose of our discussion, the main feature of this Hamiltonian is that it supports propagation of the individual fluxons (see, e.g. \cite{sol}):
\begin{equation}
\chi(x,t) =4 \tan^{-1} [\exp((x-vt)/ \lambda_0 )] \, .\label{e4}
\end{equation}
This expression describes the fluxon propagating along the array with a small velocity $v\ll (L_0 C_t)^{-1/2}$. In the situation desired in the context of quantum computation, the energy losses in the array are negligible. The fluxon motion in this regime is ballistic, i.e., the velocity $v$ is determined by the process of fluxon injection into the array. In the ballistic regime, fluxon propagation can be used for measurements of superconducting qubits \cite{averin2006,herr2007,fedorov2013,fedorov2014}. If small energy losses in the dynamics of the common mode are non-negligible, velocity $v$ of the fluxon motion is established by the balance between these losses and the driving force created by the applied bias current \cite{scott1978,pedersen1984,ustinov1998}. In the present context, such a weakly-dissipative regime of the fluxon motion does not prevent fluxons from serving as carriers of quantum information, as long as dissipation is confined to the common mode dynamics. In both regimes, velocity $v$ is directly related to the dc bias voltage across the nSQUIDs of the array, which can be used to control and monitor the fluxon motion. In long nSQUID arrays we consider in this work, propagating fluxon described by Eq.~(\ref{e4}) serves as the clock pulse transporting along the array the excitations of the differential phase $\phi(x,t)$ that encodes quantum information.

\section{Localized excitations as information carriers}

The Hamiltonian $H^{(\phi)}$ (\ref{e4b}) describing the dynamics of the phase $\phi(x,t)$ is roughly similar to the Hamiltonian of an individual SQUID that are used in typical superconducting qubits. The main new feature of the phase $\phi(x,t)$ in comparison to the phase in usual qubits is that $\phi(x,t)$ is now a field with dependence on coordinate $x$, and accordingly, the eigenstates of $H^{(\phi)}$ have a spatial structure being distributed along the array. This spatial structure is controlled by the variation of the clock phase $\chi(x,t)$ in the fluxon (\ref{e4}) which modulates the energy density in $H^{(\phi)}$ (\ref{e4b}). The lowest-energy excitations of the information phase $\phi(x,t)$ are localized in the region where $\chi \simeq \pi$ and the effective Josephson coupling energy $2E_J\cos \chi$ is negative and largest in absolute value. These excitations can be used to encode information, e.g., by serving as the basis state of a qubit localized on the fluxon. As one can see from Eq.~(\ref{e4b}),  depending on the relative magnitude of the Josephson coupling strength and inductance $L$, dynamics of the phase $\phi(x,t)$ in the $\chi \simeq \pi$ region is governed by either a bi-stable potential as required for encoding the qubit of information in two different flux states in the flux qubits \cite{flux1,flux2}, or a monostable potential in which the information can be encoded in two different energy states, similarly to the phase qubits \cite{phase}. In either case, in the nSQUID array, the qubit is attached to the fluxon and moves with it along the array.

Naturally, to account for the bi-stable dynamics of $\phi$ one needs to keep higher-order terms in $\phi^2$ in the Hamiltonian $H^{(\phi)}$ (\ref{e4b}). In the following, we consider quantitatively the situation similar to the phase qubits, when the relative magnitude of the Josephson coupling is such that
\[ \beta \equiv 2E_J L/ (\hbar/2e)^2 <1 \, ,\]
and the effective potential for $\phi(x,t)$ in the Hamiltonian (\ref{e4b}) is monostable. In this case, and for $\beta$ not
too close to 1, one can use quadratic approximation, as in Eq.~(\ref{e4b}), to determine the space structure and frequencies of information-encoding excitations of the phase $\phi(x,t)$.

For the purpose of quantum information transfer, one is interested in the regime of the sufficiently slow evolution of $\chi (x,t)$, when the characteristic frequencies of the dynamics of $\phi$, which are on the order of $\omega_p=(2LC)^{-1/2}$, are much larger than the frequency associated with the fluxon propagation, $\dot{\chi} \simeq v \ll\omega_p$. In this, adiabatic,  regime, the qubit transfer process along the array that is driven by the time evolution of $\chi (x,t)$ in the moving fluxon (\ref{e4}), with exponential accuracy does not affect the states of phase $\phi$. Quantitatively, for adiabatic evolution of $\chi$, one can neglect the time dependence in Eq.~(\ref{e4}) when calculating the structure of the excitation spectrum. Then, substituting the fluxon shape (\ref{e4}) into Eq.~(\ref{e4b}), one gets the following equation of motion for the phase $\phi(x,t)$ from the resulting Hamiltonian:
\begin{equation}
(L/L_C) \phi''=\omega_p^{-2} \ddot{\phi} +\big( 1+\beta
-\frac{2\beta}{\cosh^2x/\lambda_0}\big) \phi \, . \label{e5}
\end{equation}

As a first step, we solve this equation classically by separating the time and space dependence of $\phi(x,t)$, and expressing it as a sum of different excitation modes with some coefficients $c_j$
\begin{equation}
\phi(x,t)=\sum_j c_j \phi_j (x)e^{-i\omega_jt}.
\label{e11} \end{equation}
After this substitution, Eq.~(\ref{e5}) takes the form of an exactly solvable (see, e.g., \cite{LL}) stationary Schr\"{o}dinger equation that determines the spatial profile $\phi_j(x)$ and frequencies $\omega_j$ of the excitation modes:
\begin{equation}
-(L/L_C) \phi''-\frac{2\beta}{\cosh^2x/\lambda_0} \phi = -\big[ 1+\beta-(\omega_j/\omega_p)^2\big] \phi \, , \label{e6}
\end{equation}

We are interested in the discrete part of the spectrum of this system  which consists of the modes localized on the fluxon, in the $\chi \simeq \pi$, i.e. $x\simeq 0$, region. The qualitative features of Eq.~(\ref{e6}) as Schr\"{o}dinger equation: potential that has the  minimum value $-2\beta$ at $x=0$ and approaches $0$ at $x\rightarrow \infty$, imply that there is a continuous spectrum of frequencies of delocalized modes at $\omega \geq \omega_p (1+\beta)^{1/2}$, while the  frequencies of the modes localized on the fluxon lie within the range
\begin{equation}
\omega_p (1-\beta)^{1/2} \leq \omega_j \leq \omega_p (1+\beta)^{1/2}.
\label{e7} \end{equation}
Substituting into Eq.~(\ref{e6}) the profile of the lowest-frequency mode
\begin{equation}
\phi_0(x) = A_0 \big[ \cosh(x/\lambda_0)\big]^{-\nu_0},
\label{e8} \end{equation}
one sees that this equation is satisfied for
\[ \nu_0 = \frac{1}{2} \big[ (1+2/l)^{1/2}-1\big]\, , \]
where $l\equiv L_0/4L_C$, and gives the frequency
\begin{equation}
\omega_0 = \omega_p \big[ 1+\beta(1-\delta_0)\big]^{1/2},  \;\;\; \delta_0= \big[ (2+l)^{1/2}-l^{1/2}\big]^2 .
\label{e9} \end{equation}
For the subsequent quantization, it is convenient to have the mode profile normalized by the condition
\begin{equation}
\int dx [\phi_0(x)]^2 = 1\, .
\label{e10} \end{equation}
This condition determines the normalization constant for the zeroth mode (\ref{e8}) as
\[ A_0 = \frac{1}{\pi^{1/4}} \Big[ \frac{\Gamma(\nu_0+1/2)}{\lambda_0 \Gamma(\nu_0)}\Big]^{1/2}\, . \]
Depending on the inductances ratio $l$, the frequency (\ref{e9}) spans the whole interval (\ref{e7}). For $l \rightarrow 0$ (i.e., $L_0\ll L_C$), $\omega_0 \rightarrow \omega_p (1-\beta)^{1/2}$, qualitatively because $\nu_0$ is large and the mode is strongly localized at $x\simeq 0$, where effective Josephson coupling reaches minimum, $-E_J$. For large $l$, the mode is weakly localized, $\nu_0 \rightarrow 0$, and the frequency $\omega_0$ approaches the edge of the continuous spectrum $\omega_p (1+\beta)^{1/2}$. The middle of the interval, $\omega_0 = \omega_p$ is achieved for $L_0=L_C$, when $\nu_0=1$.

This discussion implies that the zeroth localized mode with frequency (\ref{e9}) exists for arbitrary values of the circuit parameters. For sufficiently small inductance $L_0$ (making the characteristic width $\lambda_0$ of potential well large), Eq.~(\ref{e6}) has other localized modes with higher frequencies. For any given value of the inductance ration $l$, the $j$th localized mode exists if
\[ j < \frac{1}{2} \big[ (1+2/l)^{1/2}-1\big] \, , \]
and has the frequency
\[  \omega_j = \omega_p \big[ 1+\beta(1-\delta_j)\big]^{1/2},  \;\; \delta_j= l\big[ (1+2/l)^{1/2}-(2j+1)\big]^2 . \]

As one can see from Eqs.~(\ref{e5}) and (\ref{e11}), dynamics of each localized mode is equivalent to that of a harmonic oscillator, and can be quantized in the standard way by expressing the amplitudes of the frequency components of $\phi(x,t)$ in terms of the usual bosonic  creation/annihalation operators $a_j\, , a^{\dagger}_j$. Substituting expansion (\ref{e11}) into the Hamiltonian (\ref{e4b}) and evaluating the integral by making use of the general properties of the mode functions $\phi_j(x)$: Eq.~(\ref{e6}), orthogonality, and normalization (\ref{e10}), one transforms the Hamiltonian (\ref{e4b}) into the standard form:
\[H^{(\phi)} =\sum_j \hbar \omega_j (a^{\dagger}_ja_j+1/2) \, ,  \]
and obtains the quantum version of the classical mode expansion (\ref{e11}) of the phase field $\phi(x,t)$:
\begin{equation}
\phi(x,t) = \sum_j \big(\frac{e^2}{\hbar C \omega_j}
\big)^{1/2}  \phi_j(x)\, (a_j+a^{\dagger}_j), \label{e13a}
\end{equation}
and the charge density $Q(x,t)$ associated with the dynamics of the differential mode on the junction capacitance $C$ of the nSQUIDs:
\begin{equation}
Q(x,t)= \frac{\hbar C}{e} \dot{\phi}(x,t) = i \sum_j (\hbar C \omega_j)^{1/2}  \phi_j(x)\, (a^{\dagger}_j-a_j). \label{e13b}
\end{equation}

In principle, the sums in all these expressions should run also over the continuous part of the excitation spectrum, but for the information-encoding purposes discussed in this work, only the discrete modes that are localized on the moving fluxons are of interest.
The localized modes can be used in a variety of ways to encode quantum information. The most direct approach is to use the similarity of the considered systems with the conventional phase qubits in individual SQUIDs, and use as the two basis states of the qubit of information the ground state $|0\rangle$ of the dynamics of the differential phase $\phi$ and the first excited state of the lowest-frequency mode:
\begin{equation}
\{ |0\rangle\, ,a^{\dagger}_0 |0\rangle \} \, .
\label{e14}\end{equation}
As in the phase qubits, nonlinearity of the Josephson coupling energy in the array Hamiltonian (\ref{e3}) makes it possible then to confine dynamics of $\phi$ to the two states (\ref{e14}) producing controllable qubit of information, which now, in the nSQUID configuration, is carried along the array by propagating fluxon. In the situation without the external bias for $\phi$ considered above, the lowest-order nonlinear perturbation $V$ of the quadratic Hamiltonian (\ref{e4b}) is created by the fourth-order terms in the expansion of the Josephson energy:
\begin{equation}
V=V_0(a_0+a^{\dagger}_0)^4, \label{e15} \end{equation}
\[ V_0=\frac{E_J}{12} \big(\frac{e^2}{\hbar C \omega_0}\big)^2 \int dx \phi_0^4 (x) \big(\frac{2}{\cosh^2x/\lambda_0}-1\big) \, . \]
Here, we took into account that for the purpose of using this expression in the first-order perturbation theory, one can keep in it the expansion (\ref{e13a}) of the phase $\phi$ truncated to include  only the same zeroth mode that defines the basis states (\ref{e14}). Also, for the adiabatic fluxon motion, $v\ll \omega_p$, the time dependence of $\chi(x,t)$ was neglected making the nonlinearity $V$ (\ref{e15}) a static perturbation. The perturbation $V$ creates the first-order corrections to the harmonic oscillator energies of the excitations of the zeroth mode obtained in the quadratic approximation (\ref{e6}). These corrections make the energy gap between the two qubit basis states (\ref{e14}) different by $\delta E$ from the gap separating the upper qubit state from the next energy level. Using the mode function (\ref{e8}) to calculate the spatial integral in Eq.~(\ref{e15}), we find $\delta E$ from the standard first-order perturbation theory in $V$:
\begin{equation}
\delta E =\frac{E_J}{\sqrt{\pi} \lambda_0} \Big[ \frac{e^2 \Gamma(\nu_0+1/2)}{\hbar C \omega_0\Gamma(\nu_0)} \Big]^2 \frac{(2\nu_0-1/2) \Gamma(2\nu_0)}{\Gamma(2\nu_0+3/2)}\, .
\label{e16} \end{equation}
To give a numerical example, we take $L_0=L_C$, when $\nu_0=1$ and  $\omega_0 = \omega_p=(2LC)^{-1/2}$. Then the relative magnitude of
$\delta E$ (\ref{e16}) can be expressed as $\delta E/\hbar \omega_p =(\beta/10\lambda_0) (e^2/\hbar)  (L/2C)^{1/2}$,
and can be estimated to be on the order of few tenths of a percent for typical values of parameter. The magnitude of nonlinearity can be increased by introducing the external bias into the differential mode,   which decreases the order of perturbation from the forth-order term (\ref{e15}) to the third-order term. Finite nonlinearity is needed to operate the qubit (\ref{e14}) similarly to the phase qubits, by controlling it with RF pulses. The pulse frequency is tuned to the energy difference between the basis states (\ref{e14}) of the $\phi_0$ mode, i.e., approximately to the frequency $\omega_0$ (\ref{e9}), while nonlinearity ensures that these pulses do not drive the system to the higher excitation states, limiting its dynamics to the two basis states (\ref{e14}). In this regime, the magnitude of $\delta E$ determines the  the required qubit operation time.

\section{Decoherence properties of moving qubits}

Although the structure of the nSQUID arrays is optimized for quantum information transfer, to be potentially useful, the moving qubits supported by the arrays should at least preserve the coherence properties of the current ``static'' qubit designs, making it important to understand these properties of the nSQUID qubits. NSQUID arrays  share two main physical mechanisms of decoherence with other superconducting qubits: low-frequency, typically $1/f$, noise produced by the two-level fluctuators in the materials surrounding the qubits (see, e.g., \cite{harris2008,lanting2009,yan2012}), and electromagnetic fluctuations in the control lines of the device. Although the low-frequency noise can be expected to play an even stronger role in the nSQUID arrays because of their more complex, multilayer, structure, decoherence effects of this noise are suppressed by the mechanisms inherent in the nSQUID dynamics. Part of this suppression is due to the same mechanism as in most current superconducting qubits: making the energy difference between the qubit basis states only weakly (either linearly with small coefficient, or quadratically) dependent on the external control parameters, e.g. magnetic flux or electric charge, with their low-frequency fluctuations. In the nSQUID qubits, this is the mechanisms suppressing the decoherence by the noise in the common phase $\chi(x,t)$, i.e. fluctuations in the position or shape of the fluxon (\ref{e4}) carrying the qubit. As discussed in the previous section, the basis states of this qubit are automatically localized in the region where $\chi \simeq \pi$. For such $\chi$, $\partial \cos \chi/\partial \chi \simeq 0$, and the Hamiltonian (\ref{e4b}) that governs dynamics of the differential mode depends only quadratically on the fluctuations of $\chi$. In the regime of small fluctuations of $\chi$ relevant for quantum computation, quadratic coupling suppresses their decoherence effects.

The low-frequency noise also affects the dynamics of the information-encoding differential mode $\phi(x,t)$ directly. In the situation of the quasiclassical dynamics of the Josephson phases of the nSQUIDs that is of interest for our discussion, the dominant low-frequency noise is the noise $\Phi(x,t)$ of the magnetic flux in the nSQUIDs. One can view this noise as the fluctuating part of the flux, which provides the phase bias $\phi_e(t)$ for the differential phase $\phi$ of each individual nSQUID, and couples to this phase as in the Hamiltonian (\ref{e1}). In the continuous limit, and for strong negative coupling of inductances, this noise results then in the following perturbation term that should be added to the basic Hamiltonian (\ref{e3}):
\begin{equation}
U= -\frac{\hbar}{2eL} \int dx \phi(x,t) \Phi(x,t)\, . \label{e18}
\end{equation}
Since the magnetic flux noise is produced presumably by the microscopic two-level systems localized at the superconductor-dielectric interfaces of the nSQUID structure \cite{koch2007,desousa2007,faoro2008}, the noise fluxes are uncorrelated among different nSQUIDs \cite{correlations}. In the continuous approximation, this means that the noise $\Phi(x,t)$ is $\delta$-correlated:
\begin{equation}
\langle \Phi(x,t) \Phi(x',t')\rangle =\delta (x-x') \int \frac{d\omega}{2\pi} S_0(\omega) e^{i\omega(t-t')}\, , \label{e19}
\end{equation}
where the spectral density $S_0(\omega)$ of noise in one nSQUID should have the $1/f$ profile:
\begin{equation}
S_0(\omega) =A/|\omega| \, , \label{e20}
\end{equation}
with some low- and high-frequency cutoffs, $\omega_l$ and $\omega_h$.

In the situation of the monostable potential for the differential mode $\phi(x,t)$ ($\beta<1$), and absence of the external bias for it, $\phi_e(x,t)\equiv 0$, that we focus on in this work, perturbative corrections due to $U$ (\ref{e18}) to the energies of the qubit basis states vanish in the first order, as one can argue, e.g., from the $\phi \rightarrow -\phi$ symmetry of potential in the Hamiltonian (\ref{e3}). Therefore, similarly to the fluctuations of the common mode $\chi(x,t)$, the perturbation $U$ (\ref{e18}) associated with the fluctuations of the differential mode, produces only the second-order, quadratic, non-vanishing corrections to the qubit energies, reducing the decoherence effects of these fluctuations.

Besides this suppression of the low-frequency decoherence by reduced coupling to the noise source, which is the same as in the static qubits, qubits transported along the nSQUID arrays should exhibit another suppression mechanism resulting from the {\em dynamic spreading} of the noise spectrum due to the qubit motion. Qualitatively, if the qubit is transported with velocity $v$ along the array, where the low-frequency noise is not correlated in different nSQUIDs, the qubit sees the effective noise with the correlation time on the order of $1/v$, the time of the qubit motion between the nearest-neighbor cells of the array. This mean that the spectral density of the noise as seen by the qubit is changed from the low-frequency noise spectrum of one nSQUID into the noise distributed  uniformly over the frequency range limited by $v$. Such spreading of the low-frequency noise over a large frequency range results in the change of the coherence time of the qubit from some inhomogeneous dephasing time $t_{d}$ into homogeneous dephasing time $t\sim t_{d}^{2} v$ and can be significantly increased by the qubit motion.
Such a suppression of the low-frequency qubit decoherence is qualitatively similar to the motional narrowing of the NMR lines (see, e.g., \cite{kittel}), with the main difference that it should happen not due to random thermal motion but controlled uniform propagation of the qubit.

To make this description more quantitative, we consider specifically the qubit discussed in the previous Section, with the basis (\ref{e14}) spanned by the two lowest energy states of the lowest-frequency localized excitation of the differential mode. The qubit decoherence depends on the spectral density $S(\omega)$ of the flux noise $\Phi_n(t)$ as seen by the qubit. For the qubit (\ref{e14}) transported with velocity $v$, this noise is determined by the spatial profile $\phi_0(x)$ (\ref{e8}) of the lowest-frequency excitation of the differential mode:
\begin{equation}
\Phi_n(t) = \int dx \phi_0(x-vt) \Phi(x,t)\, . \label{e22}
\end{equation}
This expression gives the following result for the noise spectral density $S(\omega)$ seen by the qubit:
\begin{eqnarray}
S(\omega)=\int d\tau \langle \Phi_n(t+\tau) \Phi_n(t)\rangle e^{-i\omega\tau} \nonumber \\ = \int d \omega'  S_0(\omega')
f(\omega -\omega' )\, , \label{e23}
\end{eqnarray}
where the function $f(\omega)$ is defined as
\begin{equation}
f(\omega)= \frac{1}{2\pi} \int dt \int dx \phi_0(x) \phi_0(x-vt) e^{i\omega t}\, . \label{e24}
\end{equation}
Taking into account the normalization condition (\ref{e10}), one can see directly from Eq.~(\ref{e24}) that $f(\omega)$ satisfies the two basic conditions:
\begin{equation}
\int d\omega f(\omega) =1\, ,  \;\;\; f(0)= \frac{1}{2\pi v} \Big[\int dx \phi_0(x)\Big]^2 \, . \label{e27}
\end{equation}

Qualitatively, this function describes how the noise spectral density is spread over the large frequency range by qubit motion. For stationary qubit, $v=0$, normalization condition (\ref{e10}) shows   that $f(\omega -\omega')=\delta (\omega -\omega')$, and the qubit states distributed over several nSQUIDs in the array see the same noise (\ref{e23}) as in one nSQUID:
\[ S(\omega)= S_0(\omega) \, .\]
For rapid qubit motion, when the frequency range up to a large frequency proportional to $v$ encloses all intensity of the low-frequency noise, $\omega_h\ll v$, Eq.~(\ref{e23}) reduces to:
\begin{equation}
S(\omega)= Wf(\omega), \;\;\; W\equiv \int d \omega  S_0(\omega)\, . \label{e28} \end{equation}

In general, the qubit motion does not change the total noise intensity, as one can see from Eqs.~(\ref{e23}) and (\ref{e27}):
\[ \int d\omega S(\omega) =W\, , \]
but it changes it distribution over frequencies. In particular, for the rapid motion, the total intensity of the original noise $S_0(\omega)$  is distributed over the whole frequency range given by the qubit velocity, and the resulting noise spectrum $S(\omega)$  is completely determined by the motion, as in Eq.~(\ref{e28}). As one can see from the second equation in (\ref{e27}), an important consequence of this is that $S(\omega)$ becomes flat, $S(\omega)\simeq S(0)$ in the large frequency range $\omega\ll v$, and its magnitude is suppressed by velocity as $1/v$:
\[ S(0)= \frac{W}{2\pi v} \Big[\int dx \phi_0(x)\Big]^2 \, . \]
Such suppressed spectral density of effectively white noise also implies similarly suppressed qubit decoherence rate.

\begin{figure}
\includegraphics[width=8.0cm]{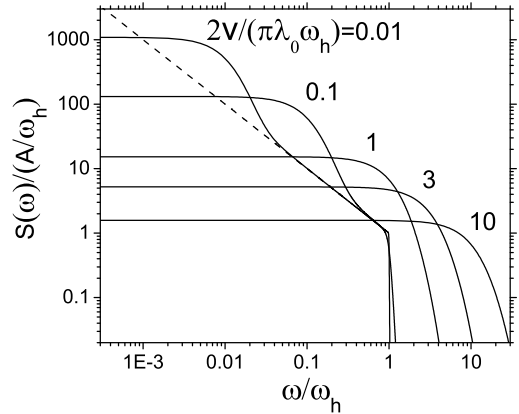}
\caption{Effective frequency spectrum of the $1/f$ noise acting on the qubit transported along the nSQUID array with velocity $v$. The noise is redistributed over frequencies by the qubit motion. The dashed line shows the original $1/f$ noise \protect (\ref{e20}). For discussion, see main text.}
\label{f3} \end{figure}
To see this more explicitly, we consider the case when the two inductances in the nSQUID array cell (Fig.~\ref{f2}) are equal, $L_C=L_0$. The spatial profile (\ref{e8}) of the lowest-frequency excitation of the differential mode then is:
\[ \phi_0(x)= [\sqrt{2\lambda_0} \cosh(x/\lambda_0)]^{-1} .\]
From this expression, one obtains explicitly $f(\omega)$ (\ref{e24}):
\begin{equation}
f(\omega)= \frac{\pi \lambda_0 }{4 v} \big[ \cosh \big(\frac{\pi \lambda_0 \omega}{2 v}\big) \big]^{-2}\, . \label{e29}
\end{equation}
Making use of this $f(\omega)$ in Eq.~(\ref{e23}) one can find the noise $S(\omega)$ for the qubit moving along the nSQUID array with an arbitrary velocity $v$, which results from the dynamic spreading of the $1/f$ noise (\ref{e20}). Figure \ref{f3} shows $S(\omega)$ obtained in this way for several velocities $v$ assuming a reasonable frequency range of the noise, $\omega_h/\omega_l=10^7$. The plot illustrates the transition from the rapid to slower qubit motion. The lowest curve (largest $v$) coincides with Eq.~(\ref{e28}) with $f(\omega)$ given by Eq.~(\ref{e29}). In agreement with the discussion above, the main feature of all curves in Fig.~\ref{f3} is that the noise spectrum becomes flat for frequencies below the characteristic frequency $\nu \equiv 2v/\pi \lambda_0$ associated with the qubit motion, inverse of the time it takes a qubit to move by a distance on the order of its  (dimensionless) ``length''. For slower qubit motion, $\omega_l \ll \nu \ll \omega_h$, the magnitude of $S$ in this low-frequency range can be estimated as
\begin{equation}
S(0)\simeq \frac{A }{\nu} \ln \big(\frac{\nu}{ \omega_l}\big) \, . \label{e30}
\end{equation}
Also, for $\nu \ll \omega_h$, the spreading function $f(\omega)$ becomes narrow, and $S(\omega)$ coincides with $S_0(\omega)$ at the higher-frequency end of the spectrum. (The sharp feature of $S(\omega)$ at $\omega =\omega_h$ in this regime is explained by the crude model of hard high-frequency cutoff used in the calculation.)

Dynamic transfer of the noise spectrum from low to high frequencies increases the qubit dephasing time. To estimate the magnitude of this increase, we assume that there is a nonvanishing differential phase bias for the nSQUIDs in the array, so that the qubit is coupled directly (linearly) to the noise, but this bias is weak and we can use quantitatively all the previous results on the noise spectrum spreading. In this case, the dephasing time $t_d$ of the stationary qubit, in the regime when it limited by the low-frequency noise producing inhomogeneous Gaussian decoherence, is $t_d=(4\pi/W)^{1/2}$ (including the relevant coupling constants into the definition of noise). The dephasing time $t_d^{(m)}$ of the same, but rapidly moving, qubit, which sees as a result the white noise spectrum producing the exponential homogeneous decoherence, is $t_d^{(m)}=2/S(0)$. Using Eqs.~(\ref{e28}) and (\ref{e29}) to find $S(0)$, we obtain the relation between the two dephasing times:
\begin{equation}
t_d^{(m)}= \nu t_d^2/\pi \, . \label{e31}
\end{equation}
From a conservative estimate of decoherence time of phase qubits, when it is dominated by the low-frequency noise, $t_{d}\sim 200$ ns, and a reasonable expected propagation frequency $\nu \sim 1$ GHz, we see that dynamic spreading of the low-frequency noise spectrum should increase the dephasing time of a qubit propagating in the nSQUID array roughly by a factor of 50, to about $t_d^{(m)}\sim 10$ $\mu$s, the value comparable to the present-day decoherence times of superconducting qubits.

\section{Summary and outlook}
In summary, we have derived the continuous model of an nSQUID array in the limit of large negative inductances of the nSQUIDs, and calculated the frequency spectrum and the spatial profile of the excitations of the differential mode in the array. Reduced to the two-state limit, dynamics of these excitations can be use to encode quantum information, which is transported along the array by propagating fluxons supported by the common mode. Qualitatively, qubit of the differential mode that is attached to the fluxon turns it into a particle with artificial spin 1/2. We have calculated the decoherence properties of these qubits which are characterized by the dynamic spreading of the noise spectrum due to qubit motion. The spreading reduces the decoherence rate associated with the low-frequency noise.

In this work, we have considered only the case of long nSQUID arrays, the common mode of which can fully accommodate individual fluxons, and made the assumption that the fluxon dynamics is classical. NSQUID arrays can be used to transport quantum information beyond this regime. The theory presented above is not directly applicable to the case of short arrays, arguably more accessible experimentally. For short arrays, the distribution of the common phase over the length of the array is different from that in a fluxon (\ref{e4}), and continuous approximation may not be valid any longer. Still, even the shortest possible ``arrays'' of two nSQUIDs should exhibit the information transfer dynamics qualitatively similar to the one considered above. Another interesting question is presented by the possibility of quantum motion of fluxons along the array, which is not completely out of the reach experimentally \cite{q_fluxon}. In this regime, when not only the differential, but also the common mode of the array behaves according to the quantum mechanics, the information propagates along the array quantum mechanically. If, in addition, the dynamics of the differential mode is bistable as in the flux qubits, the nSQUID arrays should provide then a convenient way of implementing the universal adiabatic quantum algorithms. All these possible uses of the nSQUID arrays as conveyers of quantum information merit further investigation.

\section{Acknowledgements}
The authors acknowledge useful discussions with S. Han, K.K. Likharev, and V.K. Semenov. This work was supported in part by the NSF grant PHY-1314758 and the European Union Seventh Framework Programme INFERNOS (FP7/2007-2013) under the grant agreement no.\ 308850.

\appendix*

\section{Reduction of the array Hamiltonian}

In this Appendix, we show explicitly how the Hamiltonian (\ref{e2a}) of the nSQUID array, segment of which is shown in Fig.~\ref{f2}, is reduced to the form given by Eq.~(\ref{e3a}) in the large-negative-inductance limit $M\rightarrow L$, $M_C\rightarrow L_C$. In order to do this, we start by transforming two of the initial degrees of freedom of each cell of the array, the common phase $\chi_j$ of the $j$th nSQUID, and the superconducting phase $\chi_{e,j}$ at bias node point of this nSQUID (Fig.~\ref{f2}), into two new variables $\eta_j$, $\mu_j$ defined as
\begin{equation}
\eta_j=\frac{\hbar}{2e}(\chi_j-\chi_{e,j}) \, ,\;\;\;  \mu_j=\frac{\hbar}{2eC_t}(2C\chi_j+C_0 \chi_{e,j}) \, .
\label{a1} \end{equation}
The corresponding canonical ``momenta'' conjugated to these new coordinates are the following combinations of charges $K_j$ and $K_{e,j}$:
\begin{equation}
q_j=(C_0 K_j-2C K_{e,j})/C_t \, ,\;\;\;  p_j=K_j+K_{e,j} \, .
\label{a2} \end{equation}
One can see directly that they indeed satisfy the necessary commutation relations:
\[ [\eta_j,q_j]=i\hbar \, ,\;\;  [\mu_j,p_j]=i\hbar\, , \;\;
[\eta_j,p_j]=0 \, ,\;\;  [\mu_j,q_j]=0\, . \]

This transformation is performed to separate explicitly the relative dynamics of the two phases, $\chi_j$ and $\chi_{e,j}$, which will be quenched in the limit $M\rightarrow L$, $M_C\rightarrow L_C$. To see  this, we consider the relevant, diverging in this limit, part $H'$ of the total array Hamiltonian $H_0$ [given by Eqs.~(\ref{e2a}) and (\ref{e2b})]:
\begin{eqnarray}
H'=\sum_j \Big\{ \frac{K_j^2}{4C} +\frac{K_{e,j}^2}{2C_0}  + \Big(\frac{\hbar}{2e}\Big)^2 \Big[ \frac{(\chi_{e,j} - \chi_j)^2 }{L-M} \nonumber \\ +\frac{(\chi_{e,j}-\chi_j + \chi_{j-1} - \chi_{e,j-1})^2}{L_C-M_C} \Big] \Big\} \, .
\label{a3} \end{eqnarray}
To make the subsequent calculation explicit, we need to specify the array structure. We assume the array of $N$ cells with periodic boundary conditions at the ends, e.g., $\eta_0\equiv \eta_N$. Then, inverting the relations (\ref{a2}):
\[ K_j=q_j+2C p_j/C_t\, ,\;\;\; K_{e,j}=(C_0/C_t)p_j-q_j \, ,\]
we express Eq.~(\ref{a3}) in terms of new variables:
\begin{equation}
H'=\sum_{j=1}^N \Big[ \frac{p_j^2}{2C_t}+\frac{q_j^2}{2C_r} +  \frac{\eta_j^2}{L-M} +\frac{(\eta_j-\eta_{j-1})^2}{L_C-M_C} \Big]\, ,
\label{a4} \end{equation}
where $C_r=2C C_0/(2C+C_0)$. The last three terms in this expression form the Hamiltonian $H_d$ which governs the relative dynamics of the phases $\chi_j$ and $\chi_{e,j}$. As usual, this Hamiltonian can be diagonalized by introducing plane-wave excitations with wavevectors $k=2\pi n/N,\, n=-(N-1)/2,-(N-1)/2+1, ...(N-1)/2$ (we assume for simplicity that $N$ is odd), with corresponding creation/annihalation operators $b_k\, , b^{\dagger}_k$:
\begin{eqnarray}
\eta_j=\sum_k \Big( \frac{\hbar}{2C_r\omega_k N} \Big)^{1/2}
(b_k+b^{\dagger}_{-k})e^{ikj}, \label{a5} \\
q_j=-i \sum_k \Big( \frac{\hbar C_r\omega_k }{2N} \Big)^{1/2}
(b_k-b^{\dagger}_{-k})e^{ikj} . \label{a6}
\end{eqnarray}
In terms of these plane waves, the Hamiltonian has the usual ``phonon'' form:
\[H_d =\sum_k \hbar \omega_k (b^{\dagger}_k b_k+1/2) \, ,  \]
with the following frequency spectrum:
\begin{equation}
\omega_k = \Big[ \frac{2}{C_r} \Big(\frac{1}{L-M}+ \frac{4\sin^2(k/2)}{L_C-M_C}\Big) \Big]^{1/2},
\label{a7} \end{equation}
which satisfies the condition $\omega_k=\omega_{-k}$.

The main feature of this spectrum is that all frequencies become very large, $\omega_k \rightarrow \infty$ in the large-negative-inductance limit. This means that these phonon modes can not be excited by the dynamics of the nSQUID array which occurs at much smaller frequencies, and the modes remain in the ground state throughout the time evolution of the array. The Hamiltonian $H_d$ of the relative dynamics of the phases $\chi_j$ and $\chi_{e,j}$ effectively reduces in this regime to a constant, $H_d =\sum_k \hbar \omega_k /2$, irrelevant for the system dynamics. Then, the part $H'$ (\ref{a3}) of the array Hamiltonian considered in this Appendix reduces to
\begin{equation}
H'= \frac{1}{2C_t}\sum_{j=1}^N p_j^2 +\frac{1}{2} \sum_k \hbar \omega_k \, .
\label{a8} \end{equation}

As one can see from Eq.~(\ref{a5}), suppression of the dynamics of the $\omega_k$ modes to their ground states, with added condition $\omega_k \rightarrow \infty$, also implies that $\eta_j \rightarrow 0$. Indeed, simple estimate of the fluctuations of $\eta_j$ using Eq.~(\ref{a5}) shows that $\langle \eta_j^2\rangle <\hbar [(L-M)/8C_r]^{1/2}\rightarrow 0$ for $M\rightarrow L$,  effectively imposing the constraint $\chi_j=\chi_{e,j}$ on the system dynamics. This constraint means that the ``center-of-mass'' coordinate $\mu_j$ (\ref{a1}) of these two phases also coincides with them:
$\mu_j=(\hbar/2e)\chi_j$, and the momentum $p_j$ in the Hamiltonian
(\ref{a8}) conjugate to this coordinate acts simultaneously as the momentum associated with the common phase $\chi_j$. (The difference between the two momenta can be seen only at the large frequencies $\omega_k$ not present in the system dynamics.) Taking this into account, and inserting the part $H'$ in the form (\ref{a8}), but without the irrelevant constant, back into the total array Hamiltonian $H_0$, we obtain $H_0$ in the large-negative-inductance limit $M\rightarrow L$, $M_C\rightarrow L_C$ as given by Eq.~(\ref{e3a}).


\begin{thebibliography}{99}

\bibitem{neeley2010} M. Neeley, R.C. Bialczak, M. Lenander, E. Lucero, M. Mariantoni, A.D. O'Connell, D. Sank, H. Wang, M. Weides, J. Wenner, Y. Yin, T. Yamamoto, A.N. Cleland, and J.M. Martinis,
Nature \textbf{467}, 570 (2010).

\bibitem{sun2010} G. Sun, X. Wen, B. Mao, Y. Yu, J. Chen, P. Wu, and S. Han,
Nat.\ Commun. {\bf 1}, 51 (2010).

\bibitem{paik2011} H. Paik, D.I. Schuster, L.S. Bishop, G. Kirchmair, G. Catelani, A.P. Sears, B.R. Johnson, M.J. Reagor, L. Frunzio, L.I. Glazman, S.M. Girvin, M.H. Devoret and R.J. Schoelkopf,
Phys.\ Rev.\ Lett. {\bf 107}, 240501 (2011).

\bibitem{shankar2013} S. Shankar, M. Hatridge, Z. Leghtas, K.M. Sliwa, A. Narla, U. Vool, S.M. Girvin, L. Frunzio, M. Mirrahimi, and M.H. Devoret,
Nature {\bf 504}, 419 (2013).

\bibitem{ucsb2014} R.Barends, J.Kelly, A.Megrant, A.Veitia, D.Sank, E.Jeffrey, T.C.White, J. Mutus,    A. G. Fowler, B. Campbell, Y. Chen, Z. Chen, B. Chiaro, A. Dunsworth, C. Neill, P. O'Malley, P. Roushan, A. Vainsencher, J. Wenner, A.N. Korotkov, A.N. Cleland, and J.M.Martinis,
Nature {\bf 508}, 500 (2014).

\bibitem{nist2007} M.A. Sillanpaa, J.I. Park, and R.W. Simmonds,
Nature \textbf{449}, 438 (2007).

\bibitem{ucsb2011} M. Mariantoni, H. Wang, T. Yamamoto, M. Neeley, R. C. Bialczak, Y. Chen, M. Lenander, E. Lucero, A. D. O'Connell, D. Sank, M. Weides, J. Wenner, Y. Yin, J. Zhao, A. N. Korotkov, A. N. Cleland and J.M. Martinis,
Science {\bf 334}, 61 (2011).

\bibitem{fedorov2012} A. Fedorov, L. Steffen, M. Baur, M. P. da Silva and A. Wallraff,
Nature {\bf 481}, 170 (2012).

\bibitem{mirrahimi2014} M. Mirrahimi, Z. Leghtas, V.V. Albert, S. Touzard, R.J. Schoelkopf, L. Jiang, and M.H. Devoret,
New J.\ Phys. {\bf 16} 045014 (2014).

\bibitem{dwave2011} M.W. Johnson, M.H.S. Amin, S. Gildert, T. Lanting, F. Hamze, N. Dickson, R. Harris, A.J. Berkley, J. Johansson, P. Bunyk, E.M. Chapple, C. Enderud, J.P. Hilton, K. Karimi, E. Ladizinsky, N. Ladizinsky, T. Oh, I. Perminov, C. Rich, M.C. Thom, E. Tolkacheva, C.J.S. Truncik, S. Uchaikin, J. Wang, B. Wilson, and G. Rose
Nature {\bf 473}, 194 (2011).

\bibitem{bian2013} Z. Bian, F. Chudak, W.G. Macready, L. Clark, and F. Gaitan,
Phys.\ Rev.\ Lett. {\bf 111}, 130505 (2013).

\bibitem{nsquid1} V.K. Semenov, G.V. Danilov, and D.V. Averin,
IEEE Trans.\ Appl.\ Supecond. {\bf 13}, 938 (2003).

\bibitem{nsquid2} J. Ren, V.K. Semenov, Yu.A. Polyakov, D.V. Averin,
and J.S. Tsai,
IEEE Trans.\ Appl.\ Supecond. {\bf 19}, 961 (2009).

\bibitem{nsquid3} J. Ren and V.K. Semenov,
IEEE Trans.\ Appl.\ Supecond. {\bf 21}, 780 (2011).

\bibitem{romito} A. Romito, R. Fazio and C. Bruder,
Phys.\ Rev. B \textbf{71}, 100501 (2005).

\bibitem{lyakhov} A. Lyakhov and C. Bruder,
New\ J. Phys. \textbf{7}, 181 (2005).

\bibitem{strauch} F. W. Strauch and C. J. Williams,
Phys.\ Rev. B \textbf{78}, 094516 (2008).

\bibitem{zwick} A. Zwick, G. A. \'{A}lvarez, J. Stolze and O. Osenda,
Phys.\ Rev. A \textbf{85}, 012318 (2012).

\bibitem{bruderer} M. Bruderer, K. Franke, S. Ragg, W. Belzig and D.
Obreschkow,
Phys.\ Rev. A \textbf{85}, 022312 (2012).

\bibitem{richert} J. Richert and T. Khalil,
Physica B: Cond. Matt. \textbf{407}, 729 (2012).

\bibitem{sol} R. Rajaraman, "Solitons and Instantons", (North Holland,
N.Y., 1982).

\bibitem{averin2006} D.V. Averin, K. Rabenstein, and V.K. Semenov,
Phys.\ Rev. B {\bf 73}, 094504 (2006).

\bibitem{herr2007} A. Herr, A. Fedorov, A. Shnirman, E. Il'ichev, and G. Sch\"{o}n,
Supercond.\ Sci.\ Technol.  {\bf 20}, S450 (2007).

\bibitem{fedorov2013} K.G. Fedorov, A.V. Shcherbakova, R. Sch\"{a}fer, and A.V. Ustinov,
Appl.\ Phys.\ Lett. {\bf 102}, 132602 (2013).

\bibitem{fedorov2014} K.G. Fedorov, A.V. Shcherbakova, M.J. Wolf, D. Beckmann, and A.V. Ustinov,
Phys.\ Rev.\ Lett. {\bf 112}, 160502 (2014).

\bibitem{scott1978}  D.W. McLaughlin and A.C. Scott,
Phys.\ Rev. A {\bf 18}, 1652 (1978).

\bibitem{pedersen1984} N.F. Pedersen and D. Welner,
Phys.\ Rev. B {\bf 29}, 2551 (1984).

\bibitem{ustinov1998} A.V. Ustinov,
Physica D {\bf 123}, 315 (1998).

\bibitem{flux1} J.H. Plantenberg, P.C. de Groot, C.J.P.M. Harmans,
and J.E. Mooij,
Nature {\bf 447}, 836 (2007).

\bibitem{flux2} R. Harris, J. Johansson, A.J. Berkley, M.W. Johnson, T. Lanting, S. Han, P. Bunyk, E. Ladizinsky, T. Oh, I. Perminov, E. Tolkacheva, S. Uchaikin, E. Chapple, C. Enderud, C. Rich, M. Thom, J. Wang, B. Wilson, and G. Rose,
Phys. Rev. B {\bf 81}, 134510 (2010).

\bibitem{phase} R.C. Bialczak, M. Ansmann, M. Hofheinz, E. Lucero, M. Neeley, A.D. O'Connell, D. Sank, H. Wang, J. Wenner, M. Steffen, A.N. Cleland, and J.M. Martinis,
Nat.\ Phys. {\bf 6}, 409 (2010).

\bibitem{LL} L.D. Landau and E.M. Lifshitz, {\em Quantum mechanics}
(Pergamon, 1981), Sec.~23.

\bibitem{harris2008} R. Harris, M.W. Johnson, S. Han, A.J. Berkley, J. Johansson, P. Bunyk, E. Ladizinsky, S. Govorkov, M.C. Thom, S. Uchaikin, B. Bumble, A. Fung, A. Kaul, A. Kleinsasser, M.H.S. Amin, and D.V. Averin,
Phys.\ Rev.\ Lett. {\bf 101}, 117003 (2008).

\bibitem{lanting2009} T. Lanting, A. J. Berkley, B. Bumble, P. Bunyk, A. Fung, J. Johansson, A. Kaul, A. Kleinsasser, E. Ladizinsky, F. Maibaum, R. Harris, M.W. Johnson, E. Tolkacheva and M.H.S. Amin,
Phys. Rev. B \textbf{79}, 060509 (2009).

\bibitem{yan2012} F. Yan, J. Bylander, S. Gustavsson, F. Yoshihara, K. Harrabi, D.G. Cory, T.P. Orlando, Y. Nakamura, J.S. Tsai, and W.D. Oliver,
Phys. Rev. B {\bf 85}, 174521 (2012).

\bibitem{koch2007} R.H. Koch, D.P. DiVincenzo, and J. Clarke,
Phys.\ Rev.\ Lett. {\bf 98}, 267003 (2007).

\bibitem{desousa2007} R. de Sousa,
Phys.\ Rev. B {\bf 76}, 245306 (2007).

\bibitem{faoro2008} L. Faoro and L.B. Ioffe,
Phys.\ Rev.\ Lett. {\bf 100}, 227006 (2008).

\bibitem{correlations} F. Yoshihara, Y. Nakamura and J.S. Tsai,
Phys. Rev. B {\bf 81}, 132502 (2010).

\bibitem{kittel} Ch. Kittel, {\em Introduction to solid-state physics}, (Wiley, 1996), Ch.\ 16.

\bibitem{q_fluxon} A. Wallraff, A. Lukashenko, J. Lisenfeld, A. Kemp,
M.V. Fistul, Y. Koval, A.V. Ustinov, Nature {\bf 425}, 155 (2003).


\end{thebibliography}
\end{document}